\documentclass[10pt,preprint,aps,pra,superscriptaddress]{revtex4-1}

\usepackage[normalem]{ulem}
\usepackage{bm}
\usepackage{graphicx}
\usepackage[latin1]{inputenc}
\usepackage{textcomp}
\usepackage{mathptmx}
\usepackage{amsmath}
\usepackage[usenames]{color}

\begin{document}

\title{Exploring the thermodynamics of spin-1 $^{87}$Rb Bose Gases with synthetic magnetization}

\author{Daniel Benedicto Orenes}
\email{orenes@lens.unifi.it}
\affiliation{INO-CNR Istituto Nazionale di Ottica del CNR, Sezione di Sesto Fiorentino, I-50019 Sesto Fiorentino, Italy}
\affiliation{Midlands Ultracold Atom Research Centre, School of Physics and Astronomy, University of Birmingham, Edgbaston, Birmingham, B15 2TT, United Kingdom}
\author{Anna U. Kowalczyk}
\affiliation{Midlands Ultracold Atom Research Centre, School of Physics and Astronomy, University of Birmingham, Edgbaston, Birmingham, B15 2TT, United Kingdom}
\author{Emilia Witkowska}
\affiliation{Institute of Physics, PAS, Aleja Lotnikow 32/46, PL-02668 Warsaw, Poland}
\author{Giovanni Barontini}
\email{g.barontini@bham.ac.uk}
\affiliation{Midlands Ultracold Atom Research Centre, School of Physics and Astronomy, University of Birmingham, Edgbaston, Birmingham, B15 2TT, United Kingdom}

\date{\today}

\begin{abstract}
In this work, we study the thermodynamic properties of a spin-1 Bose gas across the Bose-Einstein condensation transition. We present the theoretical description of the thermodynamics of a trapped ideal spin-1 Bose gas and we describe the phases that can be obtained in this system as a function of the temperature and of the populations in the different spin components. We propose a simple way to realize a `synthetic magnetization' that can be used to probe the entire phase diagram while keeping the real magnetization of the system fixed. We experimentally demonstrate the use of such method to explore different phases in a sample with zero total magnetization. Our work opens up new perspectives to study isothermal quenching dynamics through different magnetic phases in spinor condensates.
\end{abstract}

\maketitle 

\section{Introduction}\label{sec:intro}

Spinor Bose gases and spinor Bose-Einstein condensates (BECs) are characterized by the fact that their constituent particles have an internal degree of freedom: their spin. For example, in alkali atoms if the total spin of the atoms is $F$ and $m_F$ denotes its magnetic quantum number, the different Zeeman states of one hyperfine manifold coexist in such systems. Concerning spinor BECs, the combination of magnetic ordering and superfluidity makes them interesting systems to study phenomena involving spontaneous symmetry breaking \cite{higbie,carsten2013}, spin superfluidity \cite{lamporesi2018}, vortex dynamics \cite{lamporesi2017}, or collective magnetic excitations \cite{dan2016}. Of particular interest are the understanding of spin dynamics and the characterization of the ground states properties of these systems, which are determined by collisional processes. Collisions between the different internal states of the atoms allow  spin-changing collisions that have been studied in detail in \cite{lett2007, lett2013, kai2006,chapman2005}. These collisions can notably be employed to generate spin squeezing \cite{chapman2012,you2002,chapman2016}, that can be used to overcome the quantum shot noise limit \cite{oberthaler,carsten2014,carsten2017,carsten2018,witmetro}. Spinor dynamics was also studied in two dimensional systems \cite{carsten2014}, and in the presence of periodic potentials and across the superfluid to Mott insulator transition \cite{kaitriangular}.

Here, we focus on spin-1 bose gases, and in particular on alkali atoms in the hyperfine $F = $1 state, where the three magnetic Zeeman substates $m_F = 1, 0, -1$ coexist. Spin-1 gases can display ferromagnetic or antiferromagnetic character depending on the sign of the spin-dependent contact interaction term $c_2 = [4 \pi \hbar^2(a_2-a_0)]/3m $, where $a_F$ are the s-wave scattering lengths for the two allowed spin collisional channels $F=0$ and $F=2$, and $m$ is the atomic mass \cite{Ho}. 
The rotational symmetry of s-wave collisional processes (provided that dipolar interactions are negligible) implies that the total magnetization of the system, defined as $M = N_{+1} - N_{-1}$
with $N_{\pm 1}$ the populations in the $m_F = \pm 1$ Zeeman substates, is a conserved quantity.
Extensive work has been done to study the phases and mean-field ground states of both ferromagnetic \cite{Ho, Machida, ketterle1998Nature,Zhang2003,stamperkurn} and antiferromagnetic \cite{ketterle1999,ketterle1999bis,gerbier2012} spin-1 condensates.

While the ground state properties of spinor Bose gases have attracted substantial interest, their finite temperature behaviour has not been investigated thoroughly. The additional internal degree of freedom makes these system richer than single component Bose gases, and a large number of different thermodynamic phases can be observed. To date, only the thermodynamics of antiferromagnetic systems has been studied, where step-wise condensation of the spin components was observed as a function of the initial magnetization and the external magnetic field \cite{gerbier}. 

In this work, we study the thermodynamics of a spin-1 Bose gas using the relevant non-interacting theory for an ideal trapped gas, extending the work presented in \cite{wit}. We classify the different magnetic phases of this system, and demonstrate that it is possible to induce a `synthetic magnetization' by exploiting a spin-dependent trapping potential. We use such method to experimentally explore the phase diagram of a symmetric polar (SP) Bose gas of $^{87}$Rb, which is characterized by zero total magnetization and equal population of the three spin states $N_{0} = N_{+1} = N_{-1}$. We show that our method can be used to realize highly magnetized condensates while keeping the total magnetization of the system to zero. 

Due to their rich phase diagram, spinor Bose gases have recently been promoted as an optimal system to study non-equilibrium dynamics. For example thermal quenches were used to cross over the BEC transition \cite{ferrari2018,chapmankibble}, and microwave dressing allowed to operate selectively on the Zeemen energy levels \cite{oberthaler2018}.  Our work opens new possibilities in performing isothermal quenches across different phases using the synthetic magnetization, a method that could be exploited in future experiments to study out-of-equilibrium physics in spinor systems. 

This paper is organized as follows: In section 2 we present the non-interacting model for a spin-1 Bose gas, highlighting the presence of up to three critical temperatures and classifying the corresponding magnetic phases. In section 3, we describe our method to experimentally generate the synthetic magnetization, we present the details of our experiment and the results on the experimental exploration of the phase diagram of the $^{87}$Rb SP Bose gas.  
Finally, section 4 is devoted to the conclusions.

\section{Theory of ideal spin-1 Bose gases}\label{sec:theory}

In this section, we present the theory describing the condensation dynamics of an $F = 1$ ideal Bose gas within the grand canonical ensemble formalism. We extend the theory presented in \cite{wit} to the more general situation when not only the total atom number and the magnetization is fixed but also the number of atoms in the $m_F = 0$ Zeeman component. We will give analytical expressions for the critical temperatures and we will classify the different magnetic phases that can be realized with this kind of systems.

Let us consider an ideal, trapped, dilute spin-1 Bose gas in the presence of a magnetic field. In the case of alkali atoms, the effect of a non-zero magnetic field $B$ along the $\hat{z}$ direction, which sets the quantization axis, can be expressed analytically through the Breit-Rabi formula \cite{Zhang2003}. The contribution to the total energy of the system can be decomposed into linear and quadratic parts $E_{Zeeman} \approx - p M \, B - q N_0 \, B^2$, with $p=g_I \mu_B$, and $q = \mu_B^2 (g_I + g_J)^2/(16 E_{hfs} ) \simeq h \times 71.75$ Hz/G$^2$ for $^{87}$Rb atoms, where $g_J$, $g_I$ are the gyromagnetic ratios of the electron and nucleus, $E_{hfs}$ is the hyperfine energy splitting for zero magnetic field, $\mu_B$ is the Bohr magneton and we skipped constant terms. The linear contribution is irrelevant as it is proportional to the magnetization $M$ which is a constant of motion. The quadratic part is of the main importance in the lowest order approximation, even for a realistic system composed of interacting atoms. The presence of the spin-mixing collisional processes makes the linear part of this effect irrelevant for the dynamics. In other words, the chemical equilibrium required by the spin-changing collisional processes $|1,1\rangle + |1,-1\rangle \leftrightarrow 2 \times |1,0\rangle$ implies that the effective chemical potentials of the individual species in the condensate are constrained by the relation $\mu_1 + \mu_{-1} = 2 \mu_0$. The consequence of this is that condensing at a fixed magnetization has the same effect as condensing under the effect of an effective external magnetic field. Therefore, the applied magnetic field can be viewed as an effective magnetization of the sample.

Given these preliminary considerations, the Hamiltonian of a trapped spin-1 ideal Bose gas can be written as
\begin{equation}
H=\sum_{m_F, \vec{l}} \epsilon_{\vec{l}} n_{m_F, \vec{l}} - \mu N -\eta M-\gamma N_0 ,
\end{equation}
where $\epsilon_{\vec{l}}=l_x \hbar \omega_x+l_y \hbar \omega_y+l_z \hbar \omega_z$, $\vec{l}=(l_x,l_y,l_z)$ and $l_\alpha=0,1,2, \dots (\alpha=x,y,z)$. In the above Hamiltonian, the chemical potential $\mu$, the linear Zeeman shift $\eta$, and $\gamma$ are Lagrange multipliers that enforce all the constrains in our system i.e the conservation of the total number of atoms $N=\sum_{m_F, \vec{l}}\, \,n_{m_F, \vec{l}}$, the magnetization $M=\sum_{m_F, \vec{l}}\,\, m_F n_{m_F, \vec{l}}$, and the population of the $N_0$ state $N_0=\sum_{\vec{l}}\, \,n_{0, \vec{l}}$. The  Zeeman energy is included in the Lagrange multipliers. In other words, the Lagrange multipliers are shifted by the non-zero magnetic field, i.e. $\eta \to \tilde{\eta} + p\,B$ and $\gamma\to \tilde{\gamma} + q\,B^2$. The energy spectra for the three spin components are therefore 
\begin{subequations}
\begin{alignat}{2}
E_{1, \vec{l}} & = &&\epsilon_{\vec{l}}-\mu -\eta ,\\
E_{0, \vec{l}} & = &&\epsilon_{\vec{l}}-\mu -\gamma ,\\
E_{-1, \vec{l}}& = &&\epsilon_{\vec{l}}-\mu +\eta .
\end{alignat}
\end{subequations}

Assuming equal trapping energies $\epsilon_{\vec{l}}$ for all three Zeeman components, it is clear that a state with the lowest energy will be determined by the interplay between the external magnetic field parametrized by $\gamma$ and the effect of fixed magnetization parametrized by $\eta$. We will distinguish two limits for non-negative magnetization: {\bf a)} $\gamma > \eta$, when $E_0$ is the lowest energy state; {\bf b)} $\gamma < \eta$, when the lowest energy state is $E_1$. Using the above Hamiltonian, we can write the grand canonical partition function of the system as
\begin{equation}
\Xi = \sum_{N, M, N_0} \,\, Q_{N, M, N_0} \,\, z_\mu^N \,\, z_\eta^M \,\, z_\gamma^{N_0}, \,\,\,\,\,\, {\rm where}\,\,\,\,\,\,
Q_{N, M, N_0} = \sum_{n_{m_F, \vec{l}} } e^{-\beta E_{m_F, \vec{l}} n_{m_F, \vec{l}}}
\end{equation}
and $\beta=k_B T$ with $k_B$ being the Boltzmann constant and $T$ the temperature. The particular fugacities are $z_\mu=e^{\beta\mu}, z_\gamma=e^{\beta\gamma}, z_\eta=e^{\beta\eta} $. According to the rules of grand canonical formalism, we can derive the ensemble average population 
\begin{equation}
n_{m_F, \vec{l}}= -\frac{1}{\beta}\frac{\partial {\rm ln} \Xi}{\partial E_{m_F, \vec{l}}}
=\frac{z_{m_F}e^{-\beta E_{m_F, \vec{l}}}}{1-z_{m_F}e^{-\beta E_{m_F, \vec{l}}}},
\end{equation}
in which the effective fugacities are $z_1=z_\mu z_\eta , \, z_0=z_\mu z_\gamma, \, z_{-1}=z_{\mu} z_\eta^{-1}$. In the thermodynamic limit, following the standard derivation we can write the condensate fraction in the $m_F$ Zeeman component as
\begin{equation}
N_{m_F, 0}\equiv n_{m_F, \vec{l}=0}=\frac{z_{m_F}}{1-z_{m_F}}
\end{equation}
while the number of thermal atoms in each component can be expressed
\begin{equation}
N_{m_F,{\rm th}}\equiv \sum_{\vec{l}\ne 0} n_{m_F, \vec{l}} \cong \frac{1}{(\beta \hbar \bar{\omega})^3} g_3(z_{m_F}) ,
\end{equation} 
where $\bar{\omega}=(\omega_x \omega_y \omega_z)^{1/3}$ and $g_3(x)$ is the Bose function. 

%In fact, enforcing desired values of $N, M$ and $N_0$ is equivalent to fixing the occupations of all three Zeeman components, as $N_{\pm 1}=(N-N_0 \pm M)/2$. %Hence, one can conclude that condensation occurs in each of the component independently while the critical temperatures can be deduced a priori based on the ideal gas result for a scalar condensate $
%k_B T_{m_F, c}/(\hbar \bar{\omega}) = \left(N_{m_F}/\zeta(3)\right)^{1/3}$ where $\zeta(3)$ is the Euler-Riemann zeta function. We will provide the result by considering the two cases with special attention to the magnetization.

\begin{figure}
	\centering
		\includegraphics[width=0.6\textwidth]{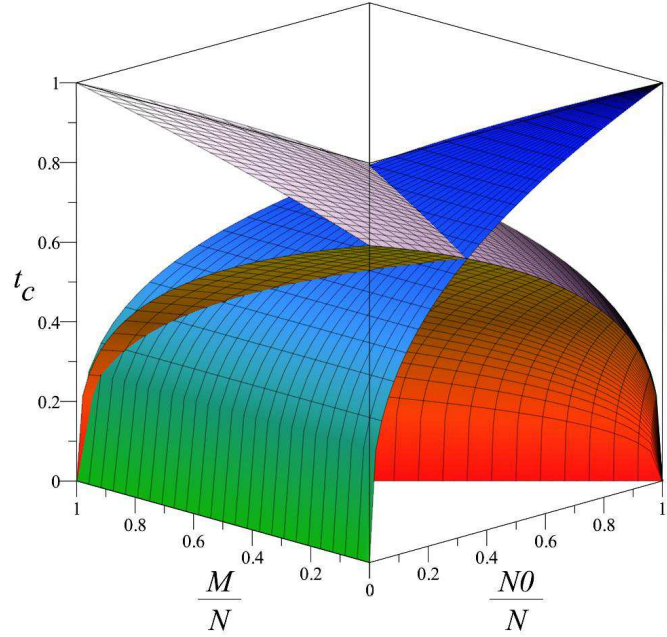}
	\caption{Normalized critical temperatures $t_0$ (orange-red), $t_1$ (blue-green) $t_{-1}$ (purple-white) as a function of the normalized population in the $m_F = 0$ state and the normalized total magnetization $M/N=(N_{+1}-N_{-1})/N$.}\label{fig:Fig1}
\end{figure}

\subsection{In the large magnetic field limit when $\gamma > \eta$}

First, we will focus on the limit of $\gamma > \eta$, i.e, when the energy associated to the quadratic Zeeman effect dominates over the mean-field energy associated to the fixed magnetization. In  this regime it is the $m_F=0$ component that condenses first, leading mathematically to $\mu \to - \gamma$, $z_0\to 1$ and $z_1\to z_\gamma^{-1} z_\eta$, $z_{-1}\to z_{\gamma}^{-1}z_{\eta}^{-1}$. The number of thermal atoms in each Zeeman state reads 
\begin{subequations}
\begin{alignat}{2}
N_{0, {\rm th}}(T) & = & \frac{\zeta(3)}{(\beta \hbar \bar{\omega})^3}, \\
N_{1, {\rm th}}(T) & = & \frac{g_3(z_\gamma^{-1} z_\eta)}{(\beta \hbar \bar{\omega})^3},\\
N_{-1, {\rm th}}(T) & = & \frac{g_3(z_\gamma^{-1} z_\eta^{-1})}{(\beta \hbar \bar{\omega})^3},
\end{alignat}
\end{subequations}
where $\zeta(3)$ is the Euler-Riemann zeta function and the number of condensed atoms are $N_{\pm	 1,0} \to 0$ and  $N_{0, 0}(T) =N_0 - N_{0, {\rm th}}(T)$. Following the arguments in \cite{wit}, the first critical temperature $T_{0, c}$ for the $m_F = 0$ state is defined by $N_0=N_{0, {\rm th}}(T_{0, c})$ what gives 
\begin{equation}
\frac{k_B T_{0, c}}{\hbar \bar{\omega}} = \left(\frac{N_0}{\zeta(3)} \right)^{1/3},
\end{equation}
because the number of atoms in the $m_F=0$ component is fixed.

The second phase transition occurs when $\eta \to \gamma$. At this point it is the $m_F=1$ component which condenses, leading to the following relations for the number of thermal atoms:
\begin{subequations}
\begin{alignat}{2}
N_{0, {\rm th}}(T) &=& \frac{\zeta(3)}{(\beta \hbar \bar{\omega})^3},\\
N_{1, {\rm th}}(T) &=& \frac{\zeta(3)}{(\beta \hbar \bar{\omega})^3},\\
N_{-1, {\rm th}}(T) &=& \frac{g_3(z_\gamma^{-2})}{(\beta \hbar \bar{\omega})^3} .
\end{alignat}
\end{subequations}
and the number of condensed atoms $N_{0,0} \gg 1$, $N_{-1, 0}\gg 1$ and $N_{1,0}(T)=N_1 - N_{1, {\rm th}}(T)$. The second critical temperature $T_{1, c}$ is defined as
\begin{equation}
\frac{k_B T_{1, c}}{\hbar \bar{\omega}} = \left(\frac{N-N_0+M}{2\zeta(3)} \right)^{1/3} .
\end{equation}

The third phase transition occurs when $\gamma \to 0$.
Assuming that $\gamma = q + \tilde{\gamma}$ as above, the transition is possible when $\tilde{\gamma}\to -q$ and then one can define the third critical temperature $T_{-1, c}$ for the $m_F = -1$ as
\begin{equation}
\frac{k_B T_{-1, c}}{\hbar \bar{\omega}} = \left(\frac{N-N_0-M}{2\zeta(3)} \right)^{1/3} .
\end{equation}

Notice, when the value of magnetization is zero then $T_{1, c}=T_{-1, c}$ and both $m_F=1$ and $m_F=-1$ components condense at the same temperature. Moreover, in the symmetric case for $N_0=N/3$ the three components condense simultaneously. In addition, the second and third phase transition can be defined also when $\eta\to - \gamma$ and then the role of $m_F=1$ and $m_F=-1$ components exchange. This case corresponds to negative values of magnetization.

\subsection{In the low magnetic field limit when $\gamma < \eta$}

\begin{figure}
	\centering
		\includegraphics[width=\textwidth]{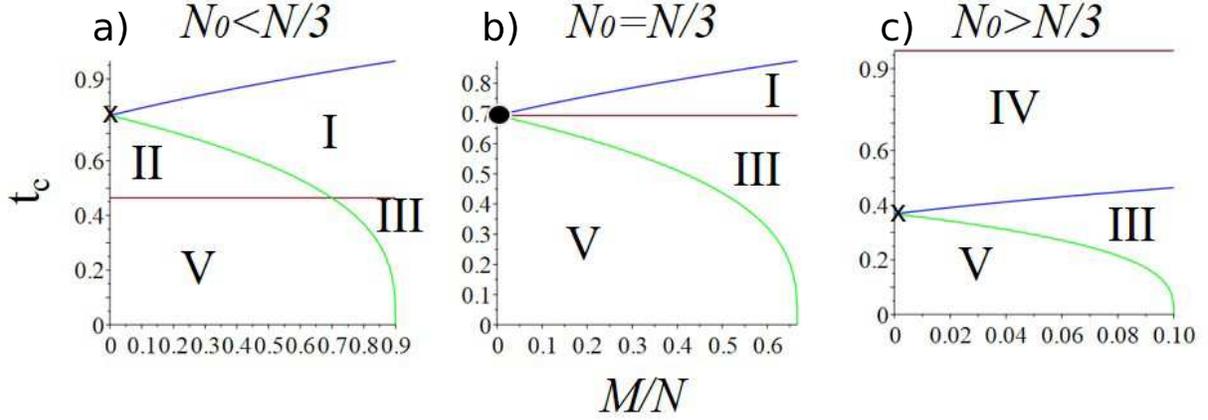}
	\caption{Normalized critical temperatures $t_0$ (red), $t_1$ (blue) and $t_{-1}$ (green) as a function of the normalized magnetization $M/N$ in the three regions: a) $\gamma>\eta$, b) $\gamma=\eta$ and c) $\gamma<\eta$. The crosses correspond to the case $M=0$ and $t_1=t_{-1}$, while the dot to the SP case, when the three critical temperatures coincide. The roman numbers indicate the different magnetic phases of the condensate: I) magnetized BEC, II) axisymmetric BEC, III) transverse magnetized BEC, IV) polar BEC and V) spinor BEC.}\label{fig:Fig2}
\end{figure}

In the case when the effect of fixed magnetization dominates it is the $m_F=1$ component that condenses first, leading to $\mu \to -\eta$ and the mathematical relations among fugacities $z_1 \to 1$, $z_0 \to z_\eta^{-1} z_\gamma$, $z_{-1}=z_\eta^{-2}$, the number of thermal atoms 
\begin{subequations}
\begin{alignat}{2}
N_{1, {\rm th}}(T)  &=& \frac{\zeta(3)}{(\beta \hbar \bar{\omega})^3}, \\
N_{0, {\rm th}}(T)  &=& \frac{g_3(z_\eta^{-1} z_\gamma)}{(\beta \hbar \bar{\omega})^3}, \\
N_{-1, {\rm th}}(T) &=& \frac{g_3(z_\eta^{-2})}{(\beta \hbar \bar{\omega})^3} ,
\end{alignat}
\end{subequations}
the condensate fractions $N_{0, 0},N_{-1,0} \to 0$ and the sharp grow of the $N_{1, 0}(T)$ value above the critical point. The first critical temperature $T_{1, c}$ for the $m_F =1$ state can be defined as
\begin{equation}
\frac{k_B T_{1, c}}{\hbar \bar{\omega}} = \left(\frac{N-N_0+M}{2\zeta(3)} \right)^{1/3}.
\end{equation}
The magnetization of condensed atoms is zero at $T=T_{1, c}$, but it starts grow up above $T_{1, c}$. The second phase transition occurs when $\eta \to \gamma$ and $N_{0,0} \gg 1$. The second critical temperature in this situation, $T_{0, c}$ can be defined from the constraint of the number of atoms in the $m_F=0$ component
\begin{equation}
\frac{k_B T_{0, c}}{\hbar \bar{\omega}} = \left(\frac{N_0}{\zeta(3)} \right)^{1/3} .
\end{equation}
At $T_{0, c}$ the magnetization of condensed atoms is already non-zero but still some thermal atoms contribute in order to take into account its fixed value
\begin{equation}
M=N_1(T_{0, c})+\frac{\zeta(3)- g_3(z_{\gamma}^{-2})}{(\beta \hbar \bar{\omega})^3}.
\end{equation}
The third phase transition takes place when $\gamma \to 0$ as we consider the limit $q\to 0$. Now, the $m_F=-1$ starts to condense. The third critical temperature $T_{-1,c}$ can be defined as 
\begin{equation}
\frac{k_B T_{-1,c}}{\hbar \bar{\omega}} = \left(\frac{N-N_0-M}{2\zeta(3)} \right)^{1/3} .
\end{equation}
What is more interesting, one can sow that at $T_{-1,c}$ the value of magnetization is determined by condensed atoms only as the contribution of thermal atoms compensate each other :
\begin{equation}
M=N_{1,0}(T_{-1,c}).
\end{equation}
An evidence of the third transition is the relative magnetization of condensed atoms equal to one, $M_c(T_{-1,c})/N_c = 1$ which is a characteristic feature of the low magnetic field region. The behaviour of the three normalized critical temperatures $t_{m_F}=T_{m_F, c}/T_c$, with $T_c=\hbar\bar\omega(N/\zeta(3))^{1/3}$ as a function of $N_0/N$ and $M/N$ is reported in Fig. \ref{fig:Fig1}. 

\subsection{Magnetic BEC phases}

From the theoretical model just described, a $F = 1$  spinor Bose gas with fixed magnetization features one, two, or three critical temperatures depending on the balance between the Zeeman populations of the sample. The presence of three different critical temperatures gives rise to a number of different phases of the condensed part of the spinor gas. In particular, for $T_{1,c}> T > T_{0,c},T_{-1,c}$ we have that only the $m_F = +1$ component is condensed, therefore we label this phase as \emph{magnetized BEC} (I). When instead $T_{1,c},T_{-1,c}> T > T_{0,c}$ both the $m_F = 1$ and $m_F = -1$ components are condensed, and we label this phase as \emph{axisymmetric BEC} (II). The case in which $T_{1,c},T_{0,c} > T > T_{-1,c}$ corresponds to a situation when the $m_F = 0$ and $m_F = 1$ components are condensed,therefore this phase corresponds to \emph{transverse magnetized BEC} (III). When $T_{0,c} > T > T_{1,c},T_{-1,c}$ only the $m_F = 0$ state is condensed and we have a \emph{polar BEC} (IV), while when $T_{1,c}, T_{0,c}, T_{-1,c} > T$ all the spin components are condensed and therefore this phase corresponds to a \emph{spinor BEC} (V). In Fig. 2 we report the different phases as a function of the total magnetization of the system for the three different cases of $\gamma > \eta$, $\gamma = \eta$ and $\gamma < \eta$. In case of zero magnetization ($\eta=0$) we have that $T_{1,c} = T_{-1,c}$. The SP state ($\eta = \gamma = 0$, indicated as a dot in Fig. 2) is the only one for which the three critical temperatures coincide and there is a direct transition from normal gas to spinor BEC.

\section{Exploring the phase diagram}

Experimentally, it is possible to access the different thermodynamic phases of a spin-1 Bose gas by adjusting the populations in the three states and changing the temperature of the sample, as it was done in \cite{gerbier}. A different method, that we employ in this work, is to use a spin-selective trapping potential that induces a `synthetic magnetization'. This allows us to control the thermodynamic properties of the system and explore the phase diagram without the need of changing the populations in the three spin components. This method opens up the possibility of performing isothermal quenches across the different thermodynamic phases, with the additional benefit that the process can be reversible. \par

Details about our experimental sequence and methods can be found in \cite{Orenes}. In brief, we load $^{87}$Rb atoms from a 3D MOT into a bichromatic crossed dipole trap made by two lasers with wavelengths $\lambda_1 = 1064$ nm and $\lambda_2 = 1550$ nm. At the beginning of the evaporation, the atoms are evenly distributed among the three Zeeman states of the $F =$ 1 ground state. This means that the total magnetization $M$ is zero (within our experimental error bars). Since no coherences are present in the system, this state is the SP = $(N/3,N/3,N/3)$ state. We then start the evaporative cooling process, that equally removes atoms from the three Zeeman states, thus preserving the magnetization and the symmetry of the state at every temperature. In other words, in our experiment we preserve the SP state at every temperature. We stop the evaporation at different times, corresponding to different final temperatures in the range $~250-0$ nK (with 0 nK we indicate a BEC with no measurable thermal component). The trapping frequencies at the end of the evaporation, i.e., when we have a pure BEC, are $2\pi\times(284; 284; 60)$ Hz. For each temperature, we let the system thermalize and equilibrate for 5 s, longer than reported in previous experimental works \cite{stamperkurn, gerbier}. This time is needed to ensure that any spin dynamics in the system has evolved towards its equilibrium state. To detect each component separately, we switch off the trapping potential and we let the atomic cloud fall during a time of flight of 30 ms. During this time, we apply a magnetic field gradient that spatially separates the three Zeeman substates. This  allows to image the three clouds independently and to fit each of them using independent routines, extracting the temperature, the number of atoms and the condensate fraction.  

\subsection{Synthetic magnetization}

As can be observed in Fig. \ref{fig:Fig2}, the SP state would normally feature a single critical temperature and the sample would undergone a one-step transition from normal gas to spinor condensate (V). However, for trapped samples, it is possible to induce a `synthetic magnetization' also for the SP state by selectively acting on the external trapping potential of the three spin components. 

Let us consider the Hamiltonian for an SP state ($\gamma=\eta=0$) with spin-selective trapping potentials: 
\begin{equation}
H=\sum_{m_F, \vec{l}} \epsilon_{m_F,\vec{l}} n_{m_F, \vec{l}} - \mu N.
\end{equation}
A way to achieve such configuration is to use a dipole trap with elliptical polarization, indeed for alkali atoms, in case of large detunings and negligible saturation, the dipole potential is given by \cite{weidemuller}
\begin{equation}
U(\textbf{r})=\frac{\pi c^2\Gamma}{2}\left(\frac{2+Pg_Fm_F}{\Delta_{2,F}\omega_2^3}+\frac{1-Pg_Fm_F}{\Delta_{1,F}\omega_1^3}\right)I(\textbf{r}), \label{eq:trapping}
\end{equation}
with $c$ the speed of light, $\Gamma$ the atom decay rate, $I(\textbf{r})$ the intensity profile of the laser, $g_F$ the Land\'e factors, $\omega_1$ and $\omega_2$  the frequencies of the $D_1$ and $D_2$ atomic transition,  $\Delta_{1,F}$ and $\Delta_{2,F}$  the corresponding detunings of the laser light and $P$ the laser polarization ($P=0$ for linear polarizations and $P=\pm1$ for $\sigma^{\pm}$ polarizations). Clearly, in case of $\sigma$ polarized light the same laser beam produces a different potential for the three spin states, while a linearly polarized or unpolarized light produces a potential that is not state dependent. In case the polarization of the light is either $\sigma^+$ or $\sigma^-$, the corresponding energy spectra are
\begin{subequations}
\begin{alignat}{3}
& E_{1, \vec{l}} &&=\epsilon_{1,\vec{l}}-\mu &&=\epsilon_{0,\vec{l}}-\mu\mp\tilde{\eta} \\
& E_{0, \vec{l}} &&=\epsilon_{0,\vec{l}}-\mu &&=\epsilon_{0,\vec{l}}-\mu \\
& E_{-1, \vec{l}} &&=\epsilon_{-1,\vec{l}}-\mu &&=\epsilon_{0,\vec{l}}-\mu\pm\tilde{\eta}  .
\end{alignat}
\end{subequations}
where it is apparent that the difference in the three trapping potentials acts as if there was a synthetic magnetization in the system. Indeed for large detunings the shift of the trapping potential for the $m_F=1$ state with respect to the potential for the $m_F=0$ state is with very good approximation opposite to the shift for the $m_F=-1$ state. Therefore, by controlling the polarization of the light from circular to linear -or not defined- we can control the synthetic magnetization and access different regions of the phase diagram. In our experiment, we achieve this by rotating the quantization axis of the system. 

\begin{figure*}[h!]
\centering
\includegraphics[width = 0.7\textwidth]{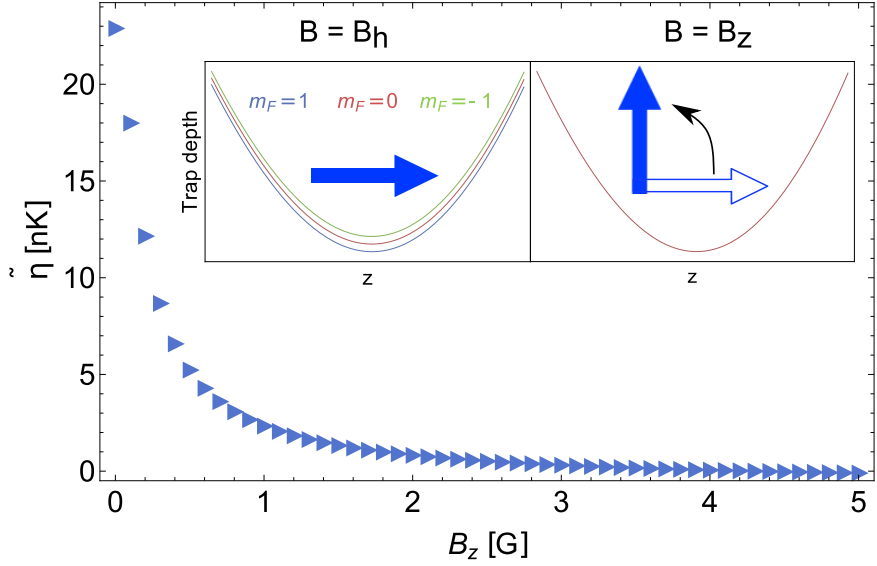}
\caption {Absolute value of the difference between the trap depth for atoms in the $m_F=\pm1$ state and atoms in the $m_F=0$ state as a function of the applied magnetic field for our trapping laser and particular experimental conditions. The insets illustrate qualitatively how the rotation of the quantization axis of the system affects the trapping potential. When the external magnetic field is directed almost vertical, the polarization of the light is no longer well defined and the trapping potential becomes spin-independent.}\label{fig:seq}
\end{figure*}

In the absence of any compensation field, in our setup there is a small horizontal magnetic field  $B_h \simeq 0.13$ G. Our trapping lasers propagate also in the horizontal plane and the one at 1064 nm have an excess of $\sigma^+$ polarization of $\simeq$ 15\%. At the beginning of the evaporation, we adiabatically ramp the current in a pair of vertical coils arranged in Helmholtz configuration, that sets the magnitude of the vertical magnetic field $B_z$. Therefore, by increasing the current in the Helmholtz coils we rotate the direction of the magnetic field. Accordingly, the quantization axis of the system rotates from horizontal to vertical (see the inset in Fig. \ref{fig:seq}), and the polarization of the light changes from elliptical to undefined. In other words, $P$ in Eq. (\ref{eq:trapping}) goes from a finite value to zero and therefore $U$ goes from being spin-dependent to be spin-independent. In Fig. \ref{fig:seq}, we plot the difference $|\Delta U_{\pm 1}| = \tilde{\eta}$ as a function of the applied magnetic field $B_z$, calculated using eq. (19) with our experimental parameters. The difference in trap depth goes from a maximum of $\simeq 25$ nK when the magnetic field is horizontal, to zero when the quantization axis is almost vertical. In our experiment, the dipole potential becomes state independent when $B_z\geq 3$ G.

\subsection{Experimental results}

\begin{figure}[h!]
	\centering
		\includegraphics[width=1\textwidth]{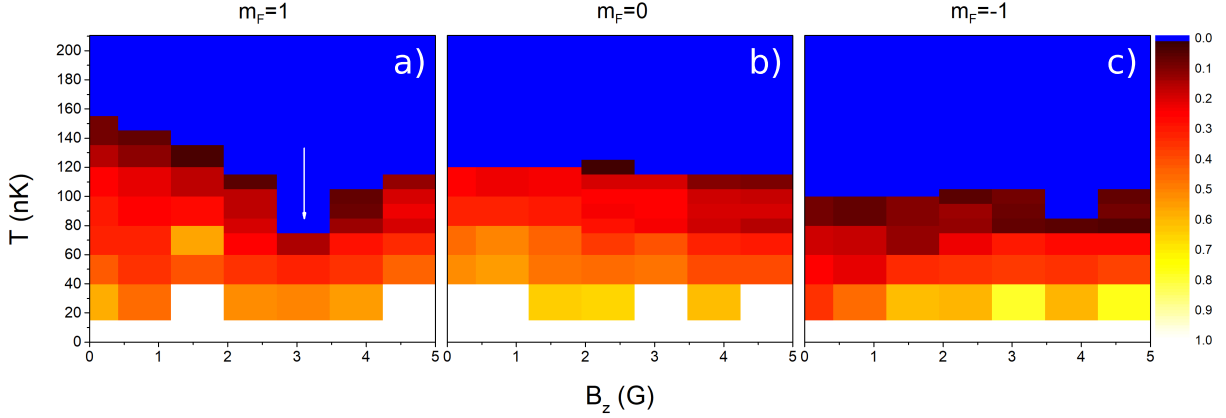}
	\caption{Measured condensate fraction for the three spin components $N_{m_F,0}$, as a function of the temperature and the applied vertical magnetic field $B_z$. Each data point corresponds to the average of at least three experimental runs. The condensate fractions are measured from independent bimodal fits performed on atomic clouds separated in time of flight by a magnetic field gradient. The white arrow indicates the value of $B_z$ for which we observe an anomalous reduction of the number of atoms in the the $m_F=1$ component.}\label{fig:condfrac}
\end{figure}

In Fig. \ref{fig:condfrac} we show the measured condensed fractions of the three spin components $N_{m_F,0}$ as a function of the temperature of the system and the applied magnetic field $B_z$. At 210 nK, which is the highest temperature shown, we have $\simeq$ 5$\times10^4$ atoms in each spin component. As we proceed with the forced evaporation towards lower temperatures, the number of atoms progressively decreases. At 50 nK we have $\simeq1.5-2\times10^4$ atoms in each spin component. We observe that for low magnetic fields the three components condense at different temperatures, while for higher values of $B_z$ the three critical temperatures coincide. At $\simeq$3 G (white arrow) we observe an anomalously low number of atoms in the $m_F=1$ component. The origin of such feature is not completely clear, however our trapping laser at 1064 nm is intrinsically modulated in amplitude at some specific frequencies. The spectrum contains a peak at $\simeq$20 MHz that at this value of the magnetic field could induce two-photon transitions to an s-wave bound state located 24.37 MHz from the atomic threshold \cite{jeremy}. The process is similar to the one studied in \cite{hutson} and its detailed study will be the subject of future works.\par 

For low values of $B_z$, where we observe three different critical temperatures, we have the highest difference between the three trapping potentials and therefore the highest value of the synthetic magnetization. As the temperature is decreased, the first atoms to condense are those in the $m_F = 1$ state, therefore realizing the \emph{magnetized BEC} phase (I). In this phase the magnetization of the condensate fraction is indeed $M_c = (N_{1,0}-N_{-1,0})/\sum N_{m_F,0}= 1$. Further decreasing the temperature, the atoms in the $m_F = 0$ condense, realizing a \emph{transverse magnetized BEC} (III), where $1 > M_c > 0$. At lower temperatures, also the atoms in the $m_F=-1$ condense and we enter the \emph{spinor condensate} phase (V). The difference between $T_{1,c}$ and $T_{0,c}$ is larger than the difference between $T_{0,c}$ and $T_{-1,c}$ even if the difference in the trapping potentials is the same. This is due to the mean-field potential exerted by the atoms already condensed that 'flattens' the potential for the non-condensed atoms, an effect not included in the non-interacting theory.    

As we increase the field, we rotate the quantization axis of the system and we therefore reduce the synthetic magnetization. We observe that the critical temperature $T_{1,c}$ progressively decreases until $T_{1,c} = T_{0,c} = T_{-1,c}$ at $\simeq 3$ G. This corresponds to the situation in which the quantization axis of the system is almost completely vertical, and the trapping potential becomes effectively spin-independent. In this regime we observe the direct transition from the normal gas to \emph{spinor BEC}. 

\begin{figure}
	\centering
		\includegraphics[width=1\textwidth]{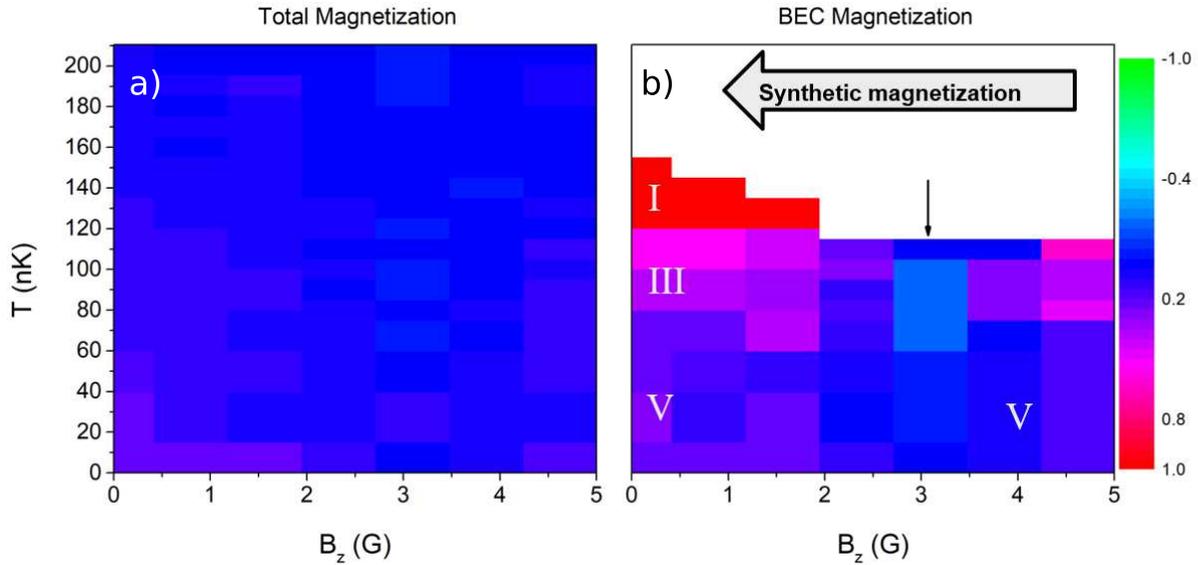}
	\caption{a) Total measured magnetization of our spinor Rb sample for different values of the temperature and of the vertical magnetic field $B_z$. b) The same but for the magnetization of the condensate fraction of the sample $M_c$. The roman numbers indicate the different regions of the phase diagram. The vertical arrow indicates the magnetic field value for which we detect lower atom number in the $m_F=1$ state.} \label{fig:phases}
\end{figure}

It is important to remark that for every value of $B_z$ the total \emph{real} magnetization of our spinor gas is always zero, as shown in Fig. \ref{fig:phases}a. Within our error bars, the system as a whole, i.e, accounting for the condensed and non condensed parts of the system, remains unmagnetized and the SP state is conserved. As discussed, the use of the spin-selective potential allows us to generate a synthetic magnetization that manifests itself in the onset of different magnetic phases for the condensed part of the sample, as shown in Fig. \ref{fig:phases}b. By changing the direction of the external magnetic field from horizontal to vertical and by controlling the temperature of the sample, we are able to explore the whole phase diagram reported in the central panel of Fig. \ref{fig:Fig2}. Increasing $B_z$ corresponds indeed to decreasing the (synthetic) magnetization and therefore to moving from right to left in Fig. \ref{fig:Fig2}b.  

\section{Conclusions and outlook}

In this work, we presented the non-interacting thermodynamic theory of a trapped spin-1 Bose gas. We classified the different magnetic phases of the condensed part of the system, and we derived analytic expressions for the critical temperatures of the three Zeeman substates $T_{m_F,c}$. We proposed a method to induce a synthetic magnetization in these systems using a spin-dependent trapping potential, and we presented a simple way to control such potential combining static magnetic fields and optical traps made with elliptically polarized light. In addition, we demonstrated experimentally that controlling the synthetic magnetization we were able to explore the phase diagram of a spin-1 Bose gas using an atomic sample with total zero magnetization. The extension of the technique presented in this work opens new exciting possibilities to study out-of-equilibrium physics in ferromagnetic spinor gases after a sudden (isothermal) quench of the synthetic magnetization, a situation that remains unexplored in the burgeoning field of quenched spinor BECs, that is attracting increasing interest both theoretically \cite{nematic,quenchdyn} and experimentally \cite{kang}.

%Experiments concerning transport and spin-dependent tunneling for the creation of highly entangled multi-particle states are highly interesting research directions.

\section*{Acknowledgments}
We are very grateful to Jeremy Hutson for performing the bound state calculations for our system. We thank J. S\'{a}nchez Claros for his help during the measurement campaign and  the people of the Cold Atoms group at the University of Birmingham for useful discussions and suggestions.

\end{document}